\documentclass[12pt]{iopart}

\usepackage{amssymb}
\usepackage{psfrag}
\usepackage{amstext}
\usepackage{epsfig}
\usepackage{longtable}
\usepackage{multirow}
\usepackage{tabularx}
\usepackage[verbose]{layout}
\usepackage{afterpage}
\usepackage{braket}

\usepackage{pstricks,pst-grad}

\begin{document}

\title{Molecular Bose-Einstein condensation in a versatile low power crossed dipole trap}

\author{J. Fuchs, G. J. Duffy, G. Veeravalli, P. Dyke, M. Bartenstein, C. J. Vale, P. Hannaford and W. J. Rowlands}

 \address{ARC Centre of Excellence for Quantum-Atom Optics, Centre for Atom Optics and Ultrafast Spectroscopy, Swinburne
University of Technology, Melbourne, Australia 3122}

\ead{jfuchs@swin.edu.au}

\begin{abstract}
We produce Bose-Einstein condensates of $^6$Li$_2$ molecules in a low power (22\,W) crossed optical dipole trap. Fermionic $^6$Li atoms are collected in a magneto-optical trap from a Zeeman slowed atomic beam, then loaded into the optical dipole trap where they are evaporatively cooled to quantum degeneracy. Our simplified system offers a high degree of flexibility in trapping geometry for studying ultracold Fermi and Bose gases.
\end{abstract}

\pacs{03.75.Ss, 32.80.Pj, 33.80.Ps}
\submitto{\JPB}
\maketitle

\section{Introduction}
Laser and evaporative cooling of neutral atoms has opened the way to studies of quantum degenerate bosonic and fermionic systems.  Recently, it has become possible to produce Bose-Einstein Condensates (BEC) of molecules, comprised of pairs of fermionic atoms, with long lifetimes~\cite{jochim03,greiner03,zwierlein03,bourdel04,partridge05,kinast05,petrov04}.  These systems are now the subject of intense investigation as they may help further our understanding of other degenerate Fermi systems such as neutron stars and high temperature superconductors.  Of particular interest is the resonant superfluid phase which occurs in the crossover region between a BEC of molecules and a  Bardeen-Cooper-Schrieffer (BCS) state of Cooper pairs~\cite{holland01}.  This has driven experimental interest and progress to date has been extremely rapid with studies of collective oscillations~\cite{bartenstein04,altmeyer07,kinast04,kinast04b}, universal behaviour~\cite{kinast05,thomas05,luo07,stewart06,partridge06}, superfluidity~\cite{chin04b,chin06,zwierlein05,zwierlein06} and polarised Fermi gases~\cite{partridge06,zwierlein06b}.

Experimentally, the generation of such Degenerate Fermi Gases (DFG) is highly complex and to date has generally relied on very high power optical dipole traps ($\sim$140\,W)~\cite{kinast05} or resonant build-up cavities~\cite{jochim03}, evaporative cooling of two different hyperfine states~\cite{greiner03} or sympathetic cooling with another isotope~\cite{bourdel04,partridge05} or species~\cite{zwierlein03,roati02,silber05}.  In this paper we describe our all-optical procedure for creating Bose-Einstein condensates of $^6$Li dimers in a lower power (22\,W) crossed dipole trap.  Our system contains a number of simplifications over previous experimental setups and allows us to easily tune the trap geometry from near spherically symmetric to highly elongated cigar shaped.

As with other $^6$Li experiments, we exploit the broad collisional Feshbach resonance centred at 834\,G~\cite{bartenstein05} to tune the interaction between particles in states $\ket{F = 1/2, m_F = +1/2}$ and $\ket{F = 1/2, m_F = -1/2}$ which we label $\ket{1}$ and $\ket{2}$, respectively.  When the magnetic field is tuned just below the resonance, where a weakly bound molecular state exists and the atomic scattering length is large and positive.  The binding energy of this weakly bound state is given by $E_B \cong \hbar^2/ma^2$ where $m$ is the atomic mass, $a$ is the $s$-wave scattering length and $\hbar$ is the reduced Planck constant~\cite{landau87}.  When the scattering length is such that the binding energy is higher than the atomic temperature (but lower than the depth of the potential confining the cloud) a trapped mixture of atoms and molecules in thermodynamic equilibrium may exist~\cite{chin04}.  As the temperature decreases below the binding energy of the molecules, the equilibrium favours molecules, which are formed by three-body recombination~\cite{jochim03b}.  This enables the production of cold molecules by the evaporative cooling of an incoherent mixture of atoms in states $\ket{1}$ and $\ket{2}$, on the $a > 0$ side of a Feshbach resonance, first demonstrated in 2003 \cite{regal03,jochim03}.

\section{Experimental Details}

\subsection{Zeeman Slower and MOT}

The starting point of our experiment is an oven of isotopically enriched $^6$Li at a temperature of 475$^\circ$C. Atoms leaving the oven are collimated by a 10\,cm long nozzle with an inner diameter of 4\,mm. The atomic beam has a mean velocity of 1900\,m/s. To obtain atoms slow enough to be captured in the Magneto-Optical Trap (MOT) the atomic beam is slowed using a Zeeman slower~\cite{phillips82}. 
The non-uniform field for the Zeeman slower is produced by a $z_0 \approx 30$\,cm long coil with an increasing magnetic field of the form $B(z) = B_0(1-\sqrt{1-z/z_0})$ with $B_0 = 620$\,G.  The magnetic field keeps the atoms at a deceleration close to $0.5a_{\mathrm{max}}$, where $a_{\mathrm{max}}$ is the atomic deceleration corresponding to full saturation of the slowing transition. 
Atoms with velocities of up to 650\,m/s are slowed down to 50\,m/s. $\sigma^-$ Zeeman slowing light is sent counter propagating to the atomic beam with a detuning of 920\,MHz with respect to the $F = 3/2$ to $F'= 5/2$ transition. 
This large detuning allows the Zeeman slowing light to pass through the MOT with no visible effect on trapping. Sidebands are added to the Zeeman slower light by modulating the laser diode current at 120\,MHz which enhances the flux of atoms trapped in the MOT by a factor of two.

The MOT is realised in a glass vacuum cell by three retro-reflected laser beams with intensities of  approximately the saturation intensity $I_\mathrm{sat}$ for both trapping and repumping. During the MOT loading phase the trapping and repumping lasers are both red detuned by 4 natural linewidths $\Gamma$ from the (unresolved) $F = 3/2$ to $F' = 1/2, 3/2, 5/2$ and $F = 1/2$ to $F' = 1/2, 3/2$ transitions, respectively.  After 40\,s loading from the Zeeman slower the MOT typically contains more than 10$^8$ atoms at a temperature of $\sim$1\,mK.  In order to maximise the number of atoms loaded into the crossed dipole trap the highest possible MOT phase space density is desired.  For this reason the MOT is compressed and further cooled for 20\,ms by increasing the magnetic field gradient from 20\,G/cm to 50\,G/cm, decreasing the detunings of both the trap and repump lasers to $\Gamma$/2 and reducing their intensities to well below $I_\mathrm{sat}$. Optical pumping into the $F = 1/2$ state is achieved by reducing the repump laser intensity more rapidly than the trapping laser.  This reduces the temperature of the cloud to 200\,$\mu$K and the rms radius to $\sim 0.5$\,mm.

The laser system used to cool the lithium atoms has been partly described in \cite{fuchs06}. Briefly, a grating stabilised external cavity diode laser is frequency stabilised by saturation spectroscopy of $^6$Li in a heated vapour cell. Four injection-locked slave diode lasers provide the laser power for trapping, repumping, Zeeman slowing and imaging.

\subsection{Optical Dipole Trap}

\begin{figure}[t]
\begin{center}
\psfrag{Yb:YAG}{\hspace{-2mm}{\small{Yb:YAG}}}
\psfrag{beam1}{\hspace{-2mm}{\small{beam dump}}}
\psfrag{beam2}{\hspace{+3mm}{\small{beam}}}
\psfrag{dump}{\hspace{+3mm}{\small{dump}}}
\psfrag{AOM}{\hspace{-1mm}{\small{AOM}}}
\psfrag{f=250mm1}{\raisebox{1mm}{\hspace{-2mm}{\small{f=250mm (L1)}}}}
\psfrag{f=250mm2}{\raisebox{1mm}{\hspace{-2mm}{\small{f=250mm}}}}
\psfrag{L3}{\raisebox{1mm}{\hspace{-2mm}{\small{(L3)}}}}
\psfrag{f=250mm3}{\raisebox{1mm}{\hspace{-2mm}{\small{f=250mm (L2)}}}}
\psfrag{lambda}{\hspace{-1mm}{\small{$\lambda$/2}}}
\psfrag{lambda2}{\hspace{-1mm}{\small{$\lambda$/2}}}
\psfrag{glasscell}{\hspace{-2mm}{\small{glass cell}}}
\psfrag{NDF}{\hspace{-2mm}{\small{NDF}}}
\psfrag{Photo diodes}{\hspace{-0mm}{\small{photodetectors}}}
\epsfig{file=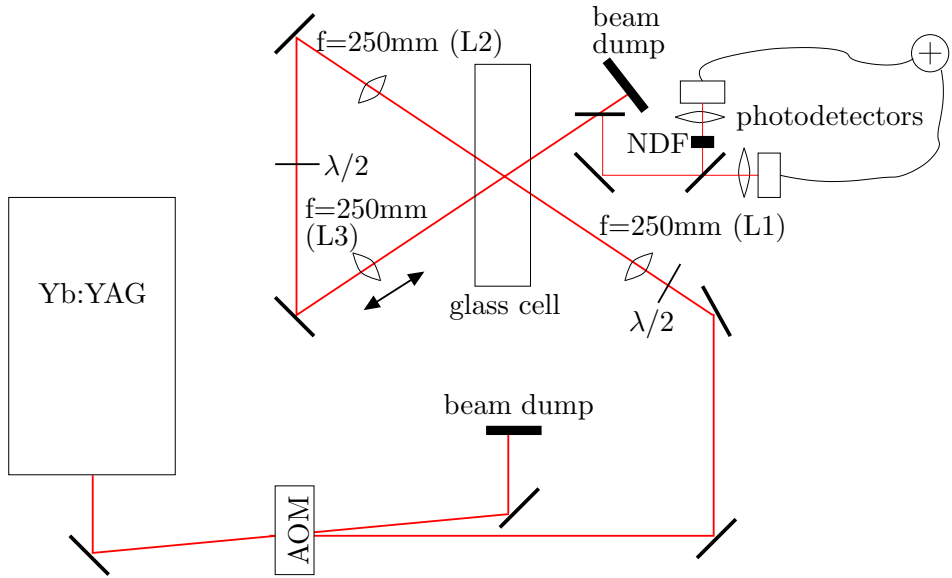,width=0.8\linewidth}
\caption{Setup of the crossed dipole trap. Before entering lens L1 the beam is near collimated with a $1/e^2$ radius of 3.4\,mm.  By translating lens L3 we can tune the aspect ratio of the trap over a wide range. }
\label{fig:setup}
\end{center}
\end{figure}

At the heart of our experiment is a crossed optical dipole trap formed by a 22\,W Yb:YAG thin disc laser (ELS VersaDisk 1030-20 SF).  Two beams of 16\,W and 14.5\,W are crossed at 80$^\circ$ at the centre of the MOT (figure \ref{fig:setup}).  To avoid interferences the polarisation of the second arm is rotated by 90 degrees with respect to the first. Both beams are focussed to a waist of approximately $30~\mu$m in the glass cell. When the beams intersect at their foci, the crossed dipole trap is nearly spherically symmetric, with an aspect ratio of 1.4. The trapping frequencies  at full power were determined by parametric driving and measurements of cloud oscillations to be $\omega_x,\omega_y,\omega_z=2\pi(16,11,11)\,s^{-1}$. The trap depths of both arms were calculated to be 0.77\,mK and 0.70\,mK, respectively.
By translating the lens used to focus the second arm along the beam we can tune the aspect ratio of the crossed dipole trap. This reduces the trap depth of the crossed part inversely to the square of the beam waist of the second arm in the overlap region. However, the total trap depth is reduced by less than 50\%.
The optical dipole trap is present during the loading and compression phases of the MOT. After extinction of the MOT beams and magnetic fields, and 1\,s of plain evaporation, up to 400,000 atoms are collected in an approximately 50/50 spin mixture of the states $\ket{1}$ and $\ket{2}$.

\begin{figure}[t]
\begin{center}
\epsfig{file=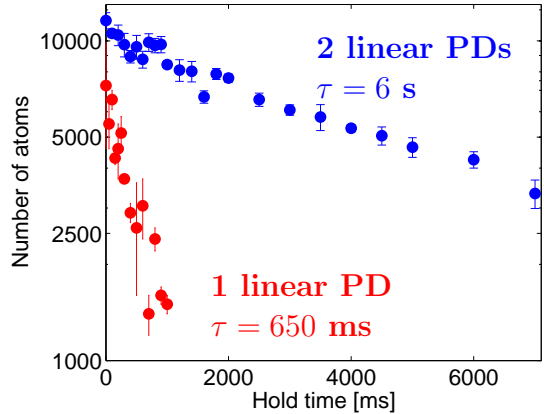,width=0.5\linewidth}
\begin{picture}(0,0)
\put(-110,135){\begin{blue} \textbf{2 linear PDs} \end{blue}}
\put(-110,120){\begin{blue} \textbf{$\tau=6$~s} \end{blue}}
\put(-150,45){\begin{red} \textbf{1 linear PD} \end{red}}
\put(-150,30){\begin{red} \textbf{$\tau=650$~ms} \end{red}}
\end{picture}
\caption{Lifetime of the crossed dipole trap with one and two photodetectors (PD) for intensity stabilisation. Number of atoms in state $\ket{1}$ for different hold times in a very shallow crossed dipole trap ($\sim$\,1\,$\mu$K). The lifetime increases from 650\,ms to approximately 6\,s due to the improved trap stability of the dual photodetector scheme.}
\label{fig:lifetime}
\end{center}
\end{figure}

To evaporatively cool these atoms it is necessary to precisely control the absolute intensity of the laser over approximately three orders of magnitude.  We achieve this by locking the laser intensity with a PID controlled Acousto-Optical Modulator (AOM).  A small fraction of the transmitted dipole trap light is measured on two photodetectors (Newfocus Model 1623).  The signals from the detectors are added together and sent to one input of the PID controller (SRS SIM 960).  This measured voltage is compared to a setpoint signal from our computer control system and used to derive an error signal which determines the amount of radio-frequency power sent to the AOM in figure \ref{fig:setup}.  Light entering one photodetector is attenuated by a factor of $\sim$\,30 using a Neutral Density Filter (NDF) so that with the same gain the signals from each detector will differ by this factor.  At high intensities, the output from the unattenuated detector is saturated and simply provides a fixed offset on the combined signal.  The signal from the other detector is below saturation and detects changes in intensity at high powers.  At lower intensities, the signal from this photodetector decreases significantly and eventually becomes negligible compared to the signal from the unattenuated detector which effectively takes over once the intensity has been reduced by a factor $\sim$\,30.  This allows us to precisely control the laser intensity over the required range for evaporation using low noise linear detectors, without the need for a logarithmic amplifier.

When only a single detector is used, the small (mV) level photodetector signals are susceptible to electronic ripple at mains frequencies which can lead to significant intensity noise on the trap laser~\cite{gehm98}. The dual photodetector scheme overcomes this problem by always allowing larger control voltages (of order a few 100\,mV) at the input to the PID.  To demonstrate this improvement, we compare the lifetime of cold clouds in very shallow traps using single and dual photodetector configurations in figure \ref{fig:lifetime}. The lifetime is seen to increase from 650\,ms to approximately 6\,s due to the improved trap stability.

\subsection{Evaporative Cooling}

\begin{figure}[t]
\begin{center}
\epsfig{file=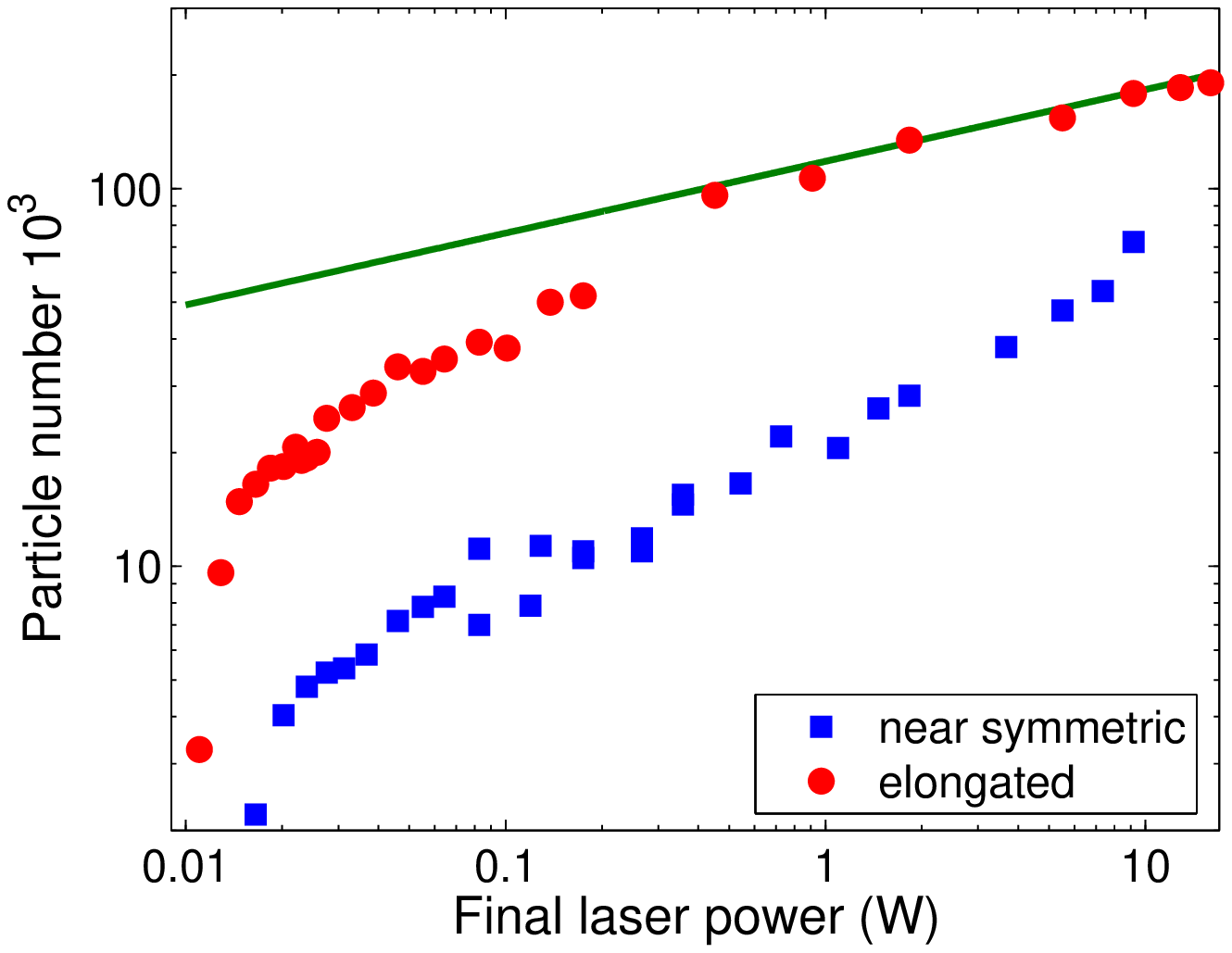,width=0.49\linewidth}
\epsfig{file=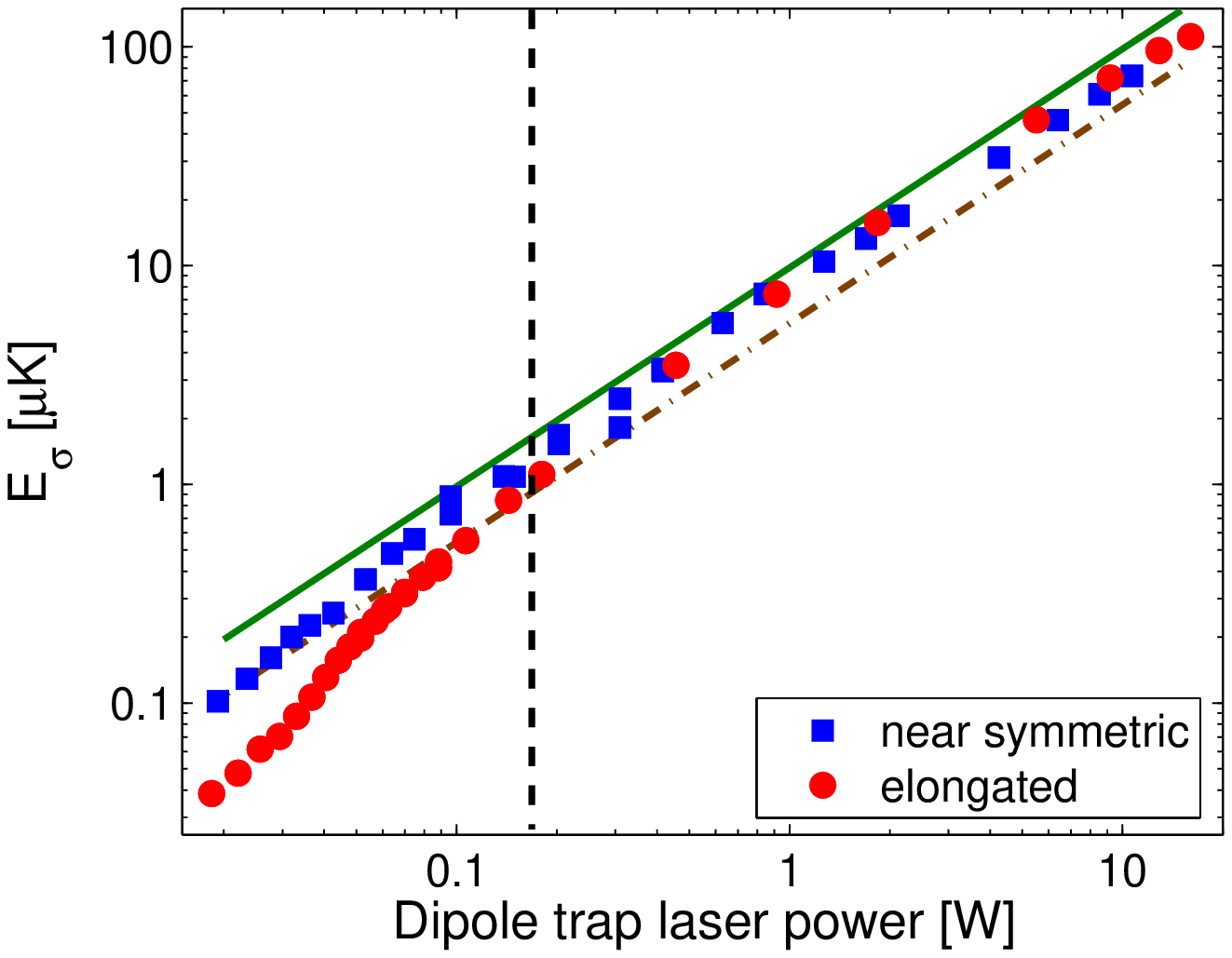,width=0.49\linewidth}
\begin{picture}(0,0)
\put(-439,160){\small{(a)}}
\put(-217,160){\small{(b)}}
\end{picture}
\caption{Evaporative cooling in the crossed dipole trap on the molecular side of the Feshbach resonance at a magnetic field of 770\,G (images taken at 694\,G). (a) Number of atoms in state $\ket{1}$ in the near symmetric (squares) and elongated (circles) crossed dipole trap is plotted versus final dipole trap laser power. The solid line represents the scaling law prediction for a cutoff parameter of $\eta=U/k_BT=10$ \cite{ohara01}. (b) Release energy as a function of final dipole trap laser power. The vertical dashed line indicates the temperature below which molecule formation becomes significant. Below the vertical dashed line the release energy no longer corresponds to the temperature of the mixture of atoms and molecules present in the trap. The solid (dash-dotted) line corresponds to one tenth of the total trap depth in the near symmetric (elongated) trapping geometry.}
\label{fig:evaporation}
\end{center}
\end{figure}

With our high trapping frequencies, tunably large scattering length and high initial phase space densities, the requirements for efficient evaporation are readily met.  A pair of high field coils (51 turns, up to 200\,A) allows us to subject the optically trapped atom cloud to magnetic fields of up to 1500\,G. This covers the entire range of the broad Feshbach resonance between the lowest two hyperfine states of $^6$Li at 834\,G.  Our evaporation consists of three stages, the first is a plain evaporation phase in which the magnetic field is turned up to 770\,G and the laser intensity is held fixed at the maximum trap depth for 1\,s.  During this stage atoms accumulate in the crossed part of the trap via elastic collisions, locally increasing the phase space density.  Next, the first ramp of the laser intensity takes place in which the laser power is lowered linearly by approximately a factor of 30 over 1.5\,s to the point where the second photodetector is no longer saturated.  Then a second sweep of 1.5\,s takes place, this time logarithmic in shape, in which the laser intensity is reduced by another factor of approximately 30. We find that the evaporation works well on either side of the Feshbach resonance when the magnitude of the scattering length is $\gtrsim 2000a_0$.

Figure \ref{fig:evaporation}(a) shows the number of atoms in state $\ket{1}$ as a function of the final laser power after evaporation at a magnetic field of 770\,G in two trapping geometries. Due to the 50/50 spin mixture of states $\ket{1}$ and $\ket{2}$ this is approximately half the total number of atoms. Absorption images are taken at 694\,G after ramping the magnetic field in 100\,ms.
The elongated trap (circles) was obtained by translating the lens L3 in figure \ref{fig:setup} by 10\,mm resulting in an aspect ratio of 15 for atoms at the bottom of the crossed dipole trap.  In this trap, the cloud is initially much hotter than the trap depth of the crossed part of the trap and a significant fraction of atoms reside in the arms, i.e. outside of the region where the beams intersect.  In the near symmetric trapping geometry (squares), however, the atoms are well confined in the crossed part after 1\,s of plain evaporation. The solid line represents the scaling laws detailed in \cite{ohara01,luo06} for an ideal evaporation with a cutoff parameter of $\eta = U/k_BT = 10$.  The atom numbers follow the scaling laws well for the first factor of 30 in the evaporation for the elongated trap.  After further evaporation, the number of atoms drops below the ideal evaporation line.  One reason for this is that the density of atoms in the arms of the crossed trap becomes very low and it is possible that not all atoms were included in the atom count.  A second reason arises from the fact that below 1\,$\mu$K a significant number of molecules will be formed (when the molecular binding exceeds the thermal energy of the cloud).  Our efficiency for detecting molecules at the imaging field is approximately 70\% of that for detecting atoms, which would account for part, but not all, of the reduction in signal \cite{jochim04}.  In the near symmetric trap (aspect ratio of 1.4), the atom number decreases more rapidly than the scaling laws predict for $\eta = 10$.  We believe this is due to the lower initial number of atoms and a  shorter trap lifetime at lower powers as this data was obtained using only a single photodetector for trap intensity stabilisation.

Also shown in figure \ref{fig:evaporation}(b) is a plot of the measured release energy of the cloud as a function of final trap depth when evaporating on the molecular BEC side of the Feshbach resonance.
The classical release energy is defined (in units of temperature) as
\begin{displaymath}
E_\sigma=\frac{m_\mathrm{atom}}{k_B}\frac{\sigma^2\omega_\mathrm{rad}^2}{1+\omega_\mathrm{rad}^2t_\mathrm{TOF}^2},
\end{displaymath}
where $m_\mathrm{atom}$ is the atomic mass, $\sigma$ is the rms radius of the cloud in the radial direction, $\omega_\mathrm{rad}/2\pi$ is the radial trapping frequency and $t_\mathrm{TOF}$ is the time of flight before taking the image. For final trap depths above 10\,$\mu$K the release energy is equal to the temperature of the atomic cloud.  Below 10\,$\mu$K a mixture of atoms and molecules will be present in the cloud and the release energy measured no longer represents the atomic temperature as the mass of the molecules is twice the mass of the atoms. Furthermore, as the molecules become degenerate the cloud profile becomes bimodal and a Gaussian fit no longer provides a representative width for a temperature measurement. The significant drop of the release energy at 50\,mW, particularly in the case of the elongated dipole trap, can be attributed to the formation of a molecular BEC, as described in section 3. 
The solid line in this figure represents one tenth of the total trap depth in the near symmetric trapping geometry, indicating that the evaporation is highly efficient with a cutoff parameter of $\eta = 10$. The dash-dotted line shows one tenth of the total trap depth in the elongated trapping configuration.  Here, the evaporation is less efficient, as at higher temperatures, a significant fraction of the atoms reside in the arms of the crossed trap where the density is lower and elastic collisions are less frequent.  As the evaporation progresses, atoms are lost from the arms and the coldest ones collect in the dimple where the two beams intersect.  As the temperature is lowered molecules form in the region of the highest atomic phase space density which is in the dimple.  The trap for molecules is twice as deep as that for atoms but, as the mass also doubles, the effective trapping frequency is the same for both.

\section{Quantum degenerate Bose and Fermi gases}

As seen in figure \ref{fig:evaporation} temperatures well below 1\,$\mu$K can be achieved for the lowest final trap powers.  When the magnetic field is tuned below the Feshbach resonance and the temperature drops below the molecular binding energy, molecules are formed and remain trapped.  The molecules can elastically scatter from each other ($a_{\mathrm{mol}} = 0.6a$ when $a$ is larger than the spatial extent of any deeply bound molecular states) and due to the low inelastic losses efficient evaporative cooling of molecules can take place~\cite{petrov04}.  These molecules are bosonic and can therefore undergo Bose-Einstein condensation at sufficiently high phase space density as has previously been reported~\cite{jochim03,greiner03,zwierlein03,bourdel04,partridge05}.

\begin{figure}[t]
\begin{center}
\epsfig{file=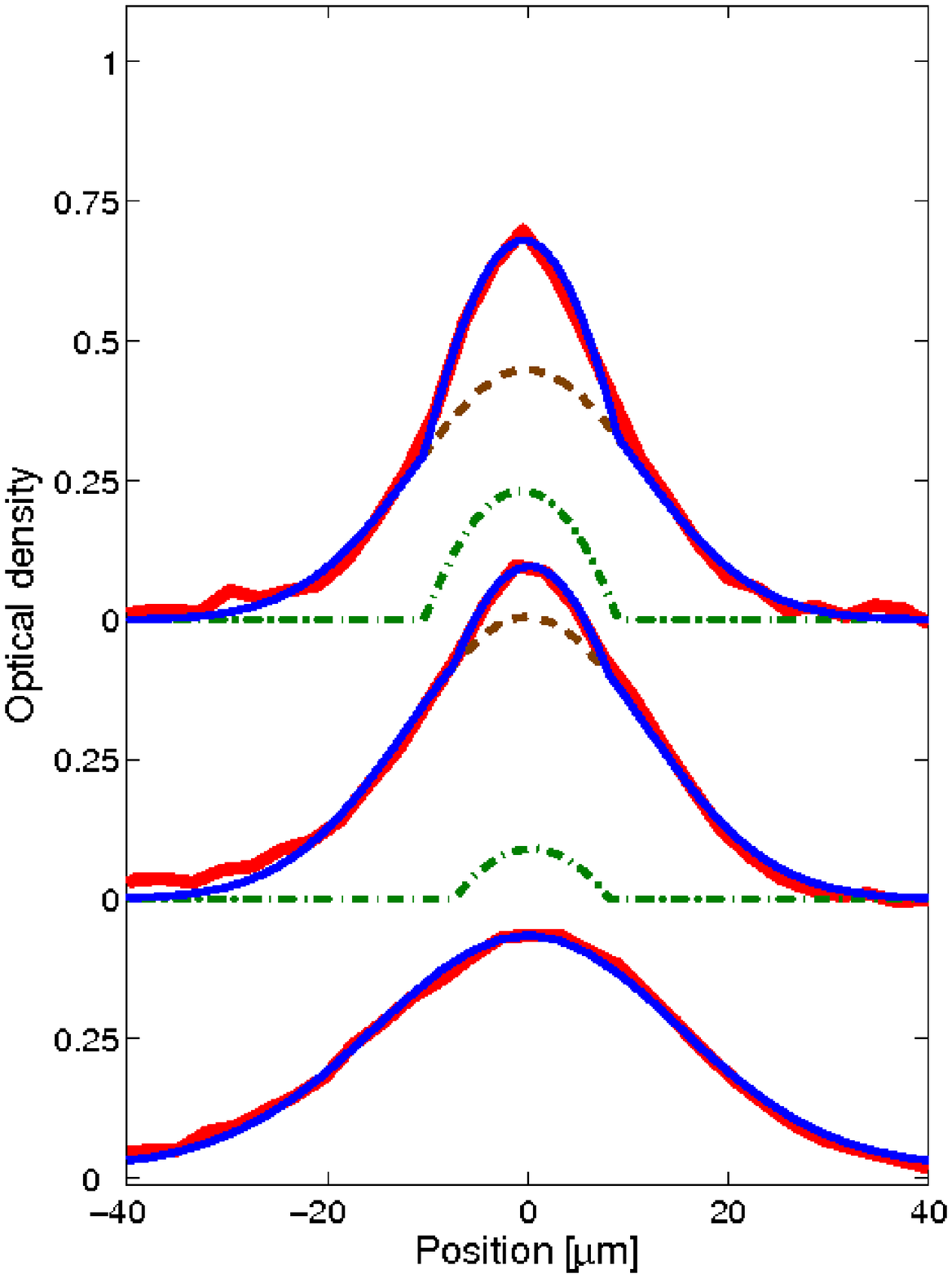,height=0.68\linewidth}
\epsfig{file=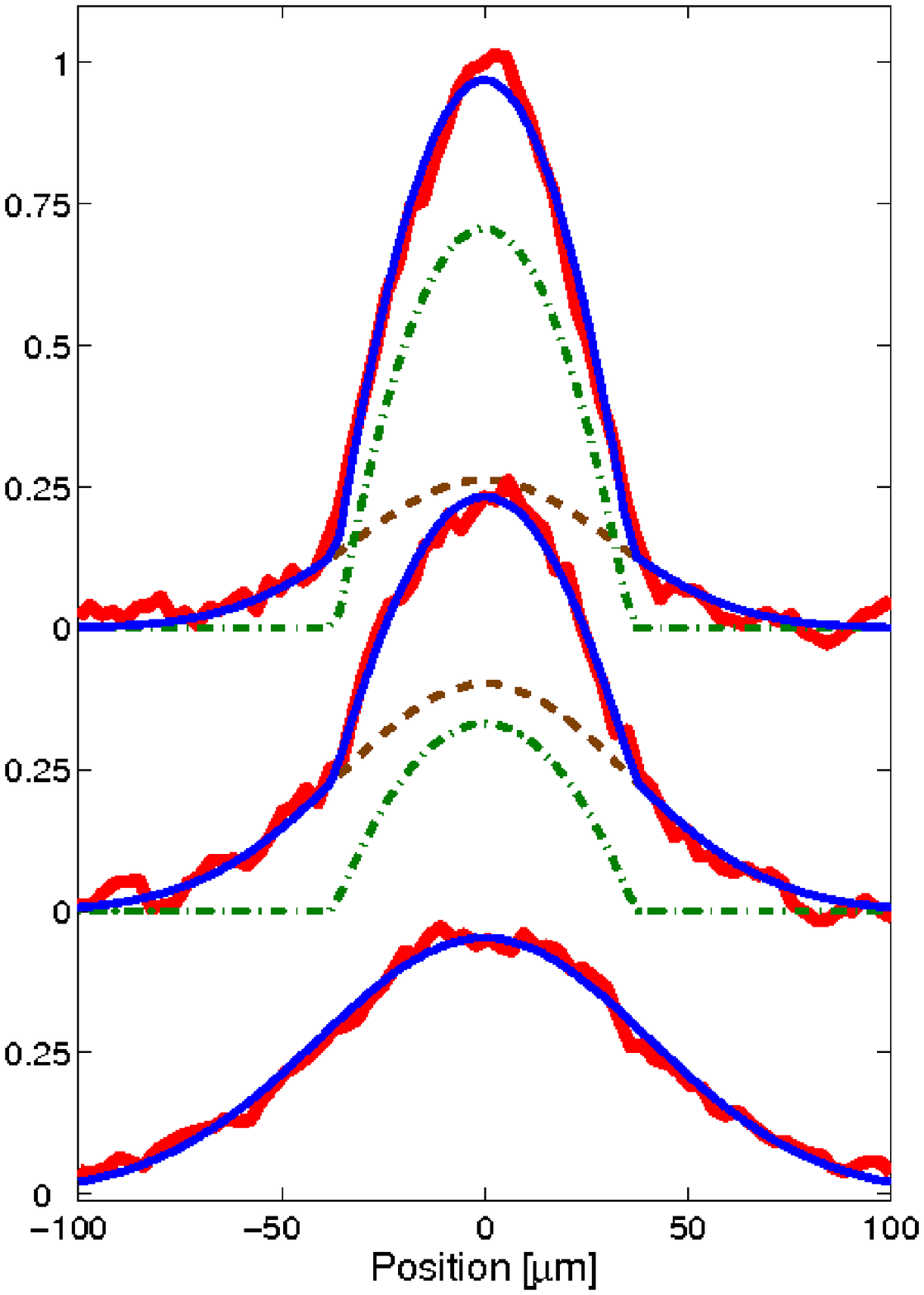,height=0.68\linewidth}
\begin{picture}(0,0)
\put(-188,145){\small{N=4000}}
\put(-188,133){\small{P=26\,mW}}

\put(-188,212){\small{N=2700}}
\put(-188,200){\small{P=15\,mW}}

\put(-188,79){\small{N=6200}}
\put(-188,67){\small{P=39\,mW}}

\put(28,145){\small{N=16000}}
\put(28,133){\small{P=25\,mW}}

\put(28,212){\small{N=12000}}
\put(28,200){\small{P=14\,mW}}

\put(28,79){\small{N=21000}}
\put(28,67){\small{P=39\,mW}}

\put(-186,305){\textbf{Near symmetric trap}}
\put(30,305){\textbf{Elongated trap}}

\end{picture}
\caption{Integrated cross sections along the weakest trapping axis 700\,$\mu$s after release from the trap showing the transition to a molecular BEC in a near symmetric (left) and elongated (right) crossed dipole trap. The dashed, dash-dotted, solid lines are fits to the Gaussian, Thomas-Fermi and combined profiles showing the thermal and condensed molecules, respectively. 
When evaporating in the elongated trap we observe a condensate fraction of 75\% for a total of 12,000 molecules. 
All images were taken at B\,=\,694\,G after evaporation at B\,=\,770\,G.}
\label{fig:bimodial}
\end{center}
\end{figure}

We have produced a molecular BEC  of  $^6$Li$_2$ in a near symmetric crossed dipole trap (with an aspect ratio of 1.4) and in an elongated trap (aspect ratio of 15).  In the elongated trap configuration we can condense up to 9,000 molecules.
This geometry also has the advantage of a smaller trapping frequency in one direction which means that the size of the BEC is much larger than our imaging resolution of $\sim$3\,$\mu$m, allowing us to discriminate the thermal and condensed molecules more easily. Absorption images are taken 700\,$\mu$s after release from the trap and the emergence of a bimodal distribution becomes evident.

Figure \ref{fig:bimodial} shows integrated cross-sections along the weakest trapping direction through expanded clouds for a near symmetric (left) and elongated crossed dipole trap (right) at various final trap depths.  As the temperature drops below the critical temperature for condensation ($\sim$250\,nK) a high density peak appears in the centre of the expanded clouds. When evaporating in the elongated trap we detect 12,000 molecules with a condensate fraction of 75\%. However, due to the reduced detection efficiency (70\%) when imaging molecules the actual number of molecules will be somewhat higher \cite{jochim04}. For the weakest elongated dipole trap shown the trapping frequencies are 23\,Hz axially and 340\,Hz radially. The dashed, dash-dotted and solid lines are fits to the Gaussian, Thomas-Fermi, and combined profiles showing the thermal and condensed fractions, respectively. The images were taken at B\,=\,694\,G after evaporation at B\,=\,770\,G.

\begin{figure}[t]
\begin{center}
\begin{minipage}{0.49\linewidth}
\epsfig{file=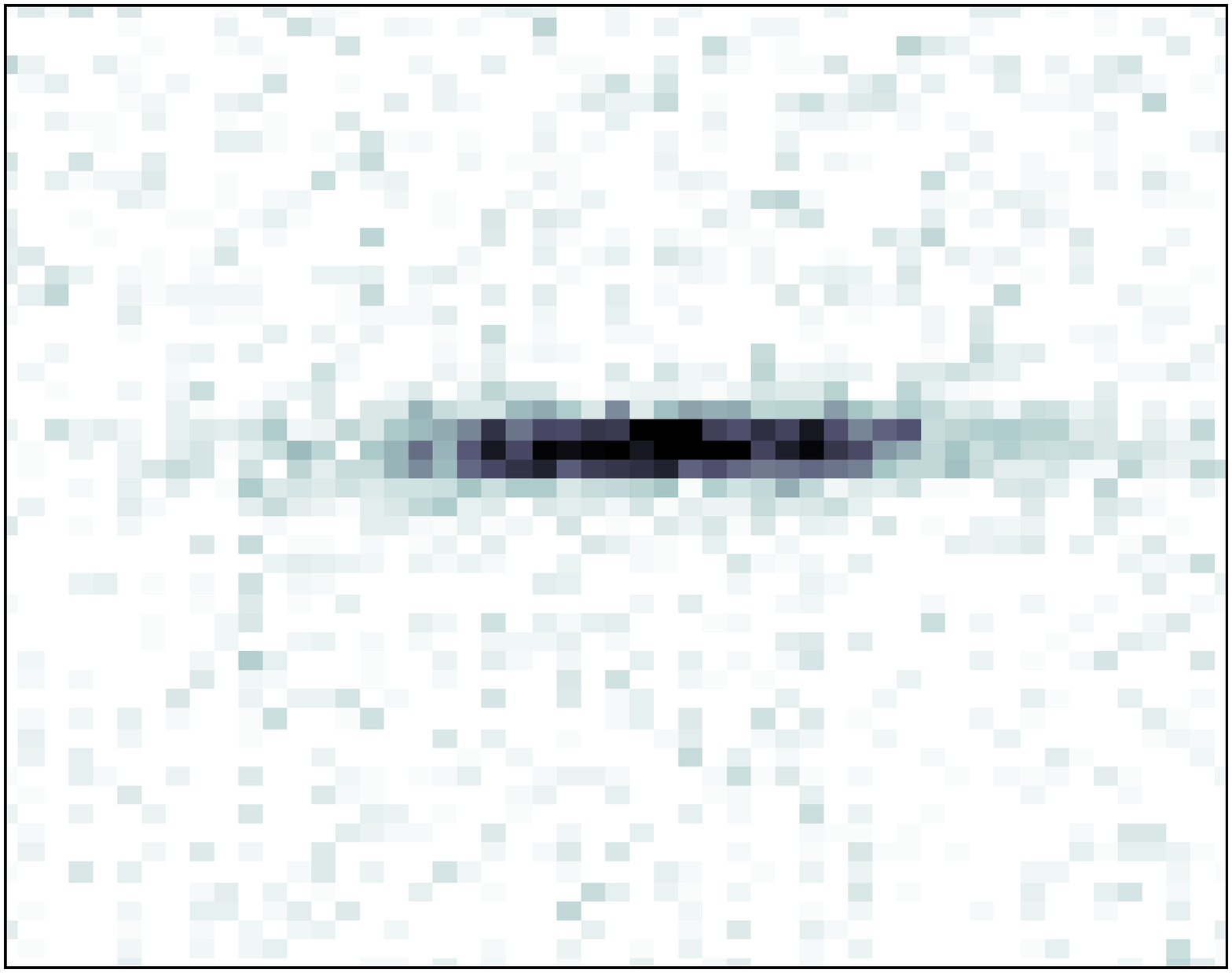,width=0.7\linewidth}\vspace{-7mm}
\epsfig{file=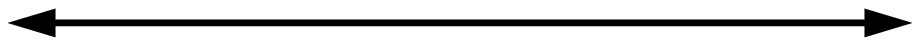,width=0.7\linewidth}
\end{minipage}
\begin{minipage}{0.49\linewidth}
\epsfig{file=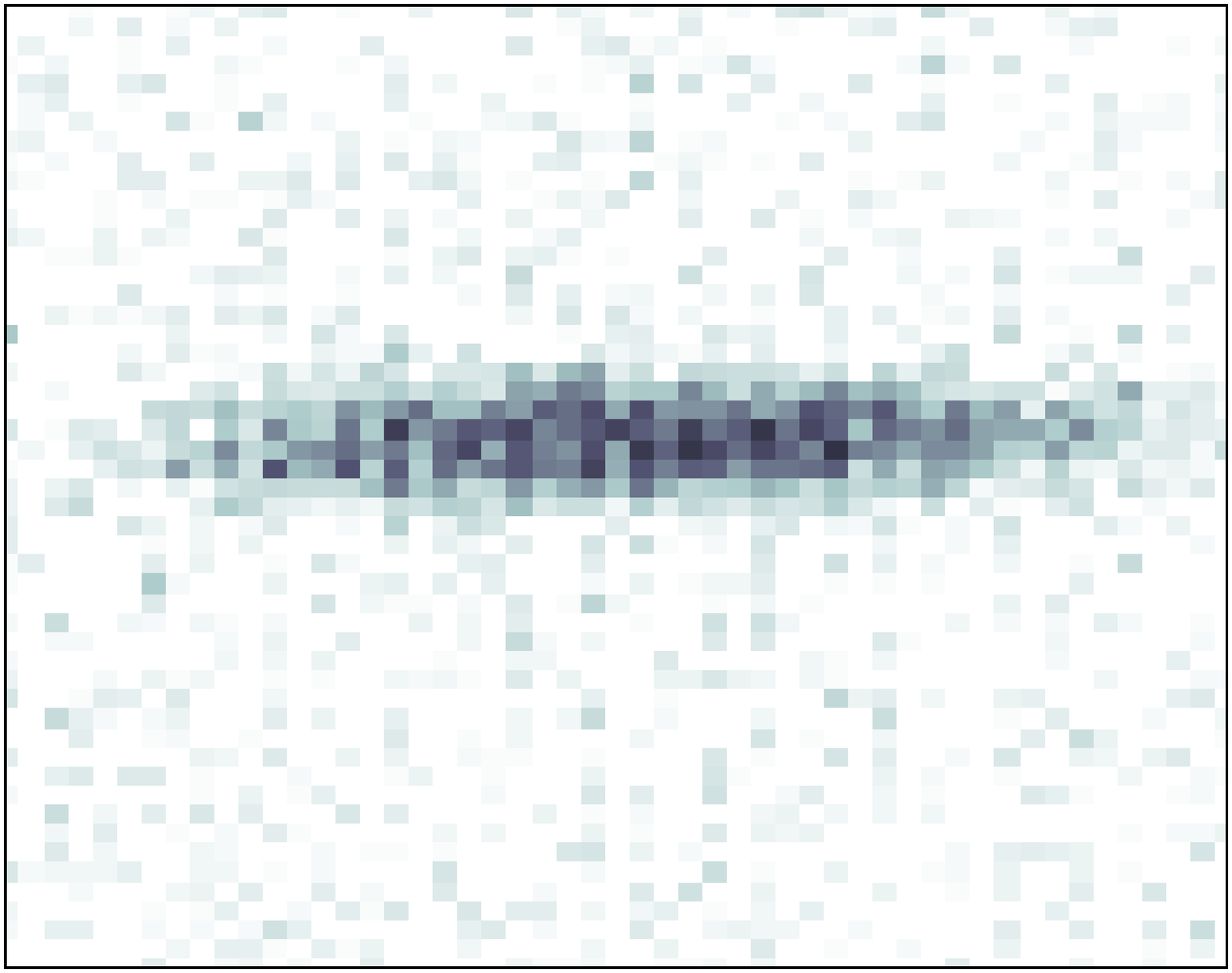,width=0.7\linewidth}\vspace{-7mm}
\epsfig{file=arrow.eps,width=0.7\linewidth}
\end{minipage}
\begin{picture}(0,0)
\put(-394,-45){150\,$\mu$m}
\put(-165,-45){150\,$\mu$m}
\put(-460,60){\small{(a)}}
\put(-238,60){\small{(b)}}
\end{picture}
\vspace{7mm}

\caption{{\it In situ} absorption images of a trapped molecular BEC and DFG, in identical crossed dipole traps following (a) evaporation at 770\,G (imaging at 694\,G) and (b) evaporation and imaging at 1100\,G. The difference between a MBEC and a DFG is clearly evident in the density and size of the clouds.}
\label{fig:BECvsBCS2}
\end{center}
\end{figure}

Degenerate bosons and fermions behave very differently and with our system we can readily compare the two cases.  A striking example is shown in figure \ref{fig:BECvsBCS2} where we compare {\it in situ} absorption images of (a) a trapped molecular BEC on the $a > 0$ (694~G) side of the Feshbach resonance and (b) a trapped degenerate Fermi gas on the $a < 0$ (1100~G) side of the Feshbach resonance.  These two cases represent two identical runs of the experiment with identical evaporation ramps (final dipole trap power 9\,mW), the only difference being the magnetic field at which the evaporation and imaging take place.  On the $a < 0$ side, there is no bound molecular state and we simply have a DFG in an incoherent spin mixture. Its distribution is wider and less dense due to the Fermi pressure.

\begin{figure}[t]
\begin{center}
\epsfig{file=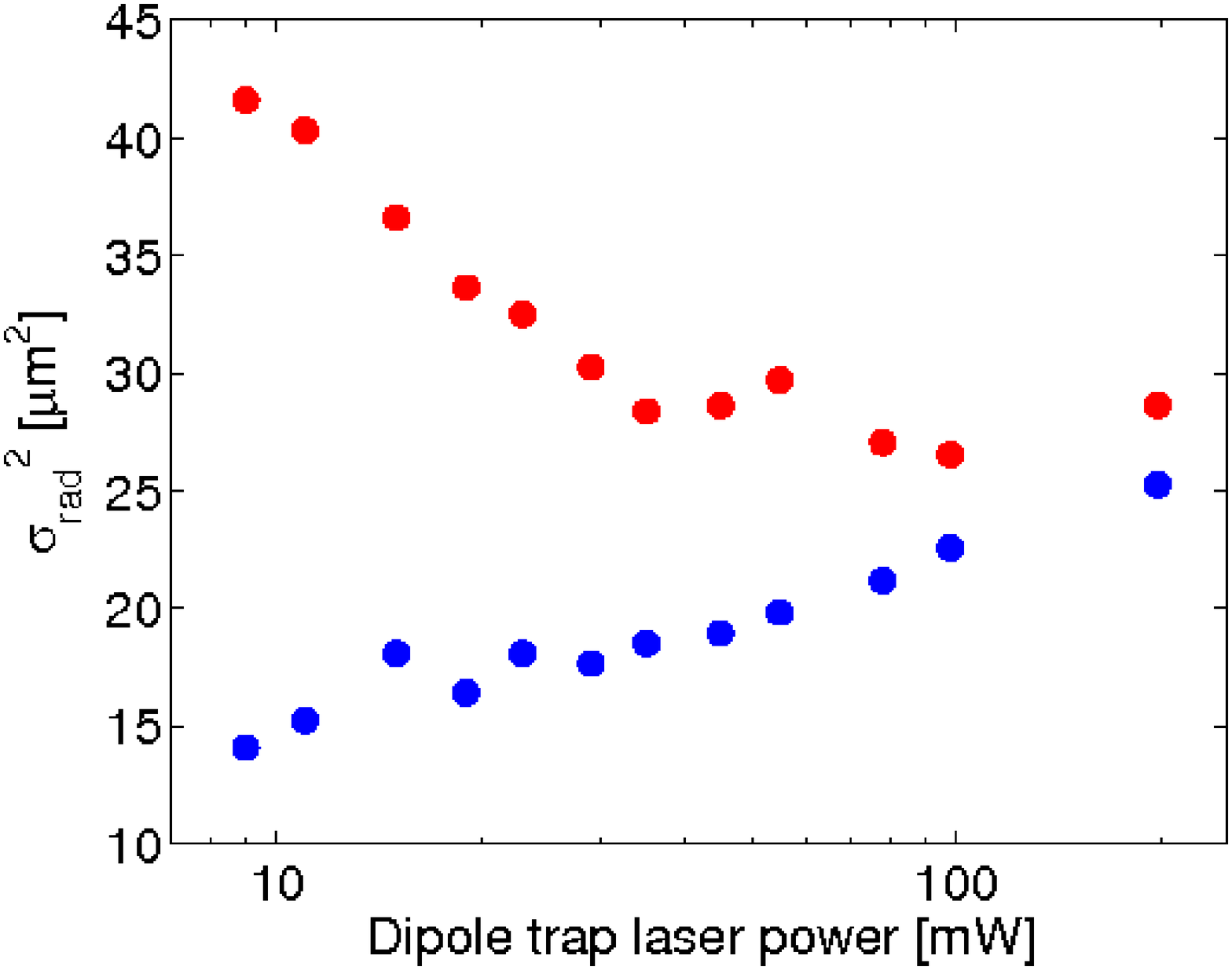,width=0.32\linewidth}
\epsfig{file=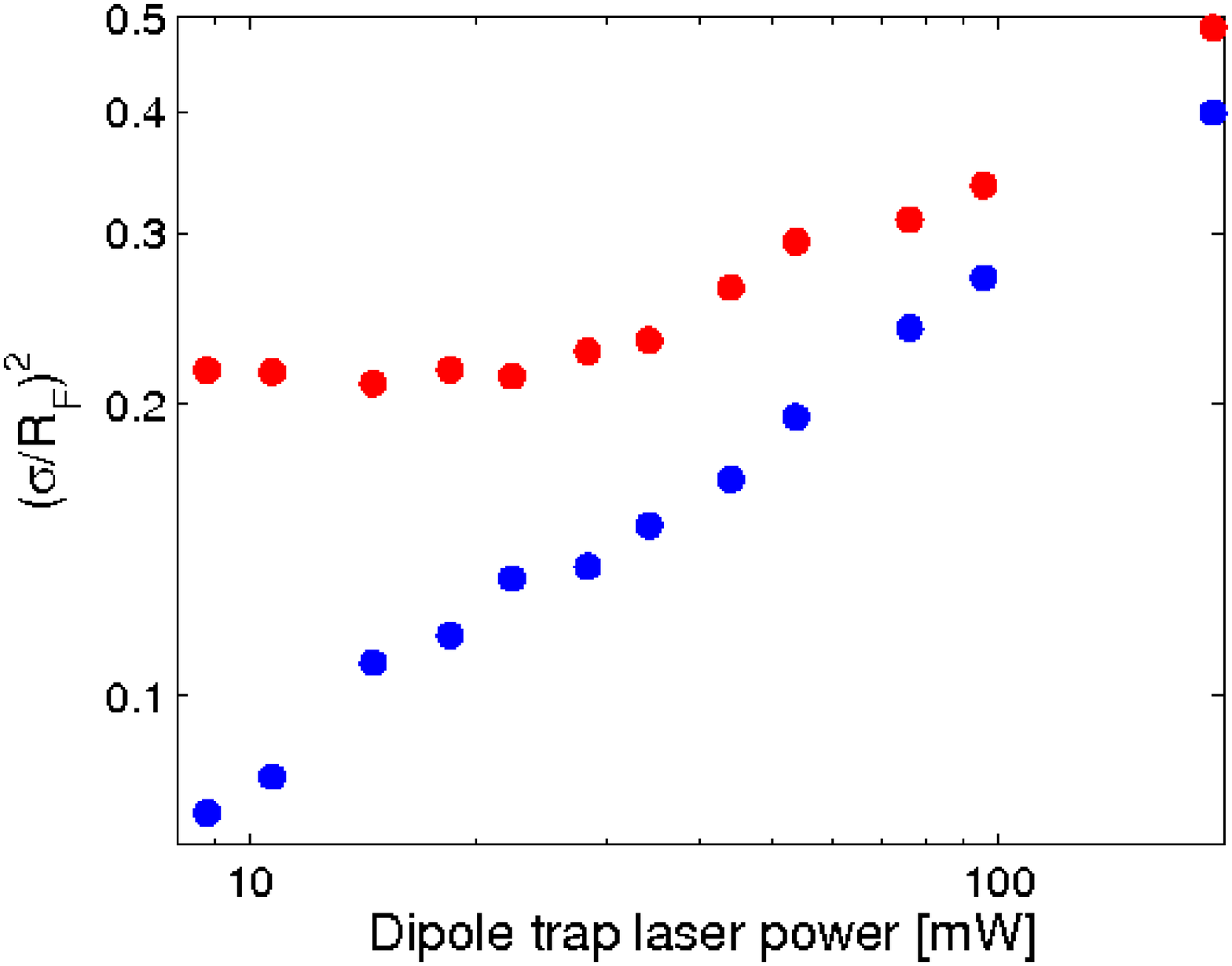,width=0.32\linewidth}
\epsfig{file=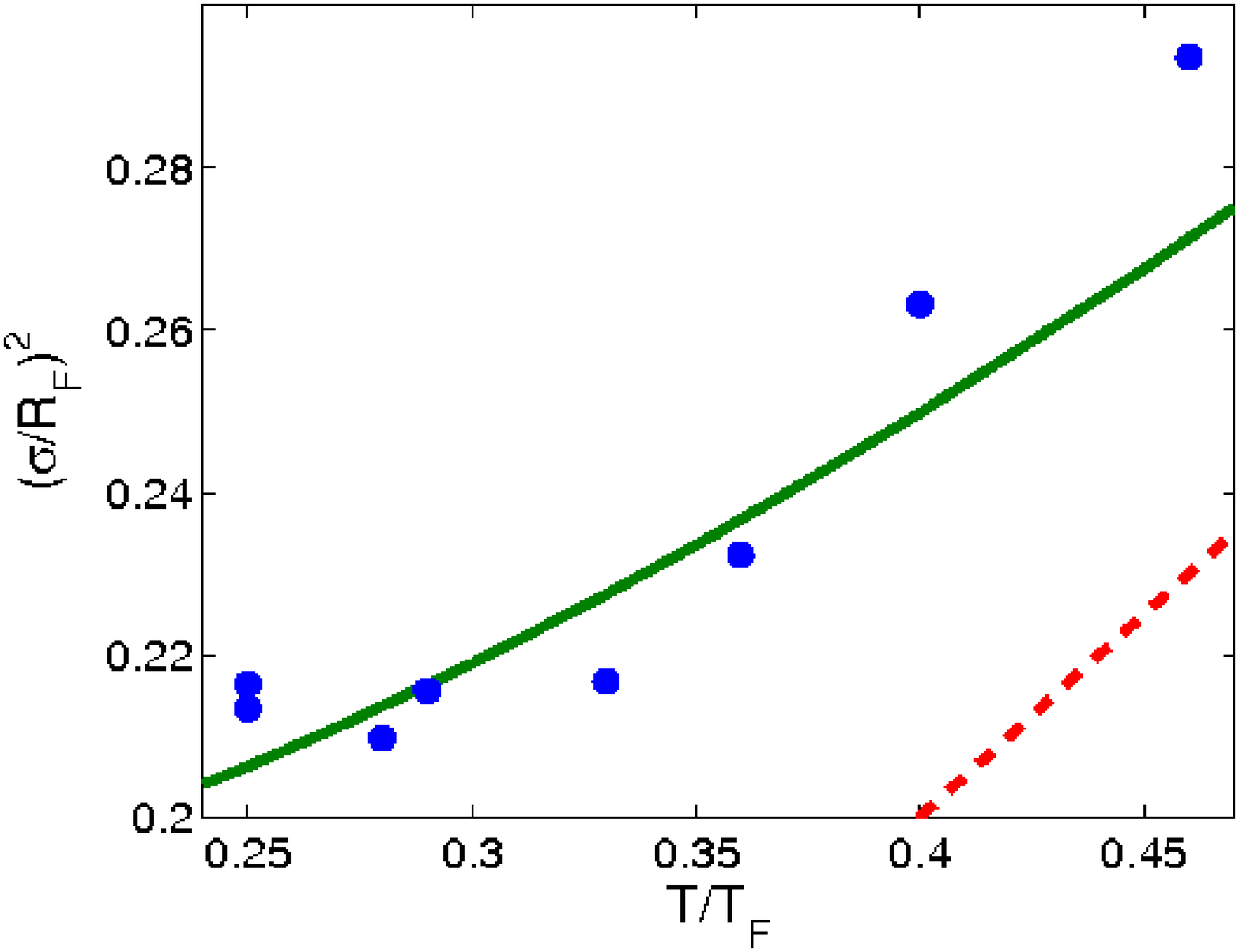,width=0.32\linewidth}
\begin{picture}(0,0)
\put(-444,110){\small{(a)}}
\put(-296,110){\small{(b)}}
\put(-142,110){\small{(c)}}
\put(-390,90){\begin{red} \textbf{$a<0$}  \end{red}}
\put(-370,20){\begin{blue} \textbf{$a>0$}  \end{blue}}
\put(-250,80){\begin{red} \textbf{$a<0$} \end{red}}
\put(-240,20){\begin{blue} \textbf{$a>0$} \end{blue}}
\end{picture}
\caption{Observation of a degenerate Fermi gas. (a) Mean square width $\sigma_\mathrm{rad}^2$ plotted versus the final dipole trap laser power on both sides of the Feshbach resonance (b) Relative mean square width ($\sigma_\mathrm{rad}/R_F)^2$ plotted versus the final dipole trap laser power (c) ($\sigma_\mathrm{rad}/R_F)^2$ plotted versus the degeneracy parameter $T/T_F$. The solid line shows the theoretical prediction without taking into account atom-atom interactions. The dashed line shows the prediction for a classical Boltzmann gas.}
\label{fig:BECvsBCS}
\end{center}
\end{figure}

The emergence of the Fermi pressure can be seen in figure \ref{fig:BECvsBCS} where we compare the widths of trapped clouds on either side of the Feshbach resonance. In figure \ref{fig:BECvsBCS}(a) we have plotted the mean square radius $\sigma_\mathrm{rad}^2$ versus the final dipole trap laser power. On the BEC side of the resonance, the distribution of bosonic molecules becomes successively narrower as the final trap depth is lowered while for fermionic atoms the width increases (the trapping frequencies reduce with the square root of the laser intensity in both cases).  Figure \ref{fig:BECvsBCS}(b) shows the same data relative to the Fermi radius ($\sigma_\mathrm{rad}/R_F)^2$ where 
$$R_F=\sqrt{\frac{2k_BT_F}{m_\mathrm{atom}\omega^2_\mathrm{rad}}}$$ 
and $T_F=\hbar \big( \omega_\mathrm{ax} \omega_\mathrm{rad}^26N\big)^{1/3}$ is the Fermi temperature.
These are also plotted against the degeneracy parameter, $T/T_F$, in figure \ref{fig:BECvsBCS}(c) at trap depths for which we have independent time of flight temperature measurements. To determine the degeneracy parameter we compared the width of Gaussian fits after 1ms time of flight to Gaussian fits to Fermi distributions. For a classical Boltzmann gas the width of the cloud decreases linearly to zero as the temperature decreases (dashed line). This also holds for fermions at the limit of high temperatures ($T\gg T_F$). However, due to the Fermi pressure the theoretical curve (solid line) vastly deviates from the classical law for low temperatures and $\sigma^2/R_F^2$  approaches 0.177 as $T$ goes to zero when integrated once over the imaging direction. The theoretical curve was determined for a non-interacting gas. However, the data in figure \ref{fig:BECvsBCS} was obtained in a spin-mixture of the two lowest hyperfine states, and due to attractive interactions on the BCS side of the Feshbach resonance the width of the cloud is expected to be slightly lower than for a noninteracting gas \cite{giorgini07}.  The highest degeneracy we have observed with our system corresponds to $T/T_F=0.25$.

\section{Summary}

We have produced Bose-Einstein condensates of $^6$Li$_2$ molecules in a low power, variable geometry optical dipole trap.  Evaporative cooling is achieved by reducing the dipole trap laser power  near the broad Feshbach resonance at 834\,G. By tuning to the low magnetic field side (770\,G) of the Feshbach resonance, molecules are formed through three-body recombination at sufficiently low temperatures. Further evaporation leads to the creation of a BEC of $^6$Li$_2$ molecules. After reducing the laser power by a factor of approximately 1000 we have observed condensates of up to 9,000 molecules. We have also produced a highly degenerate Fermi gas, as evidenced by the emergence of the Fermi pressure. Our system offers a low power dipole trap alternative for the study of degenerate Fermi gases in the BEC-BCS crossover region.

{\ack} We would like to thank Andrew Truscott for his assistance in designing the Zeeman slower,  David Lau and Selim Jochim for their contributions in the early setup stage, as well as Mark Kivinen for providing technical support and the construction of key experimental components. This project is supported by the Australian Research Council Centre of Excellence for Quantum-Atom Optics and Swinburne University of Technology. \\ \\

\end{document}